\documentclass{article}

\usepackage{arxiv}

\usepackage[utf8]{inputenc} 
\usepackage[T1]{fontenc}    
\usepackage{hyperref}       
\usepackage{url}            
\usepackage{booktabs}       
\usepackage{amsfonts}       
\usepackage{nicefrac}       
\usepackage{microtype}      
\usepackage{lipsum}		
\usepackage{graphicx}
\usepackage{natbib}
\usepackage{doi}

\title{High duty cycle EUV radiation source based on inverse Compton scattering}

\author{ \href{https://orcid.org/0000-0003-4835-7714}{\includegraphics[scale=0.01]{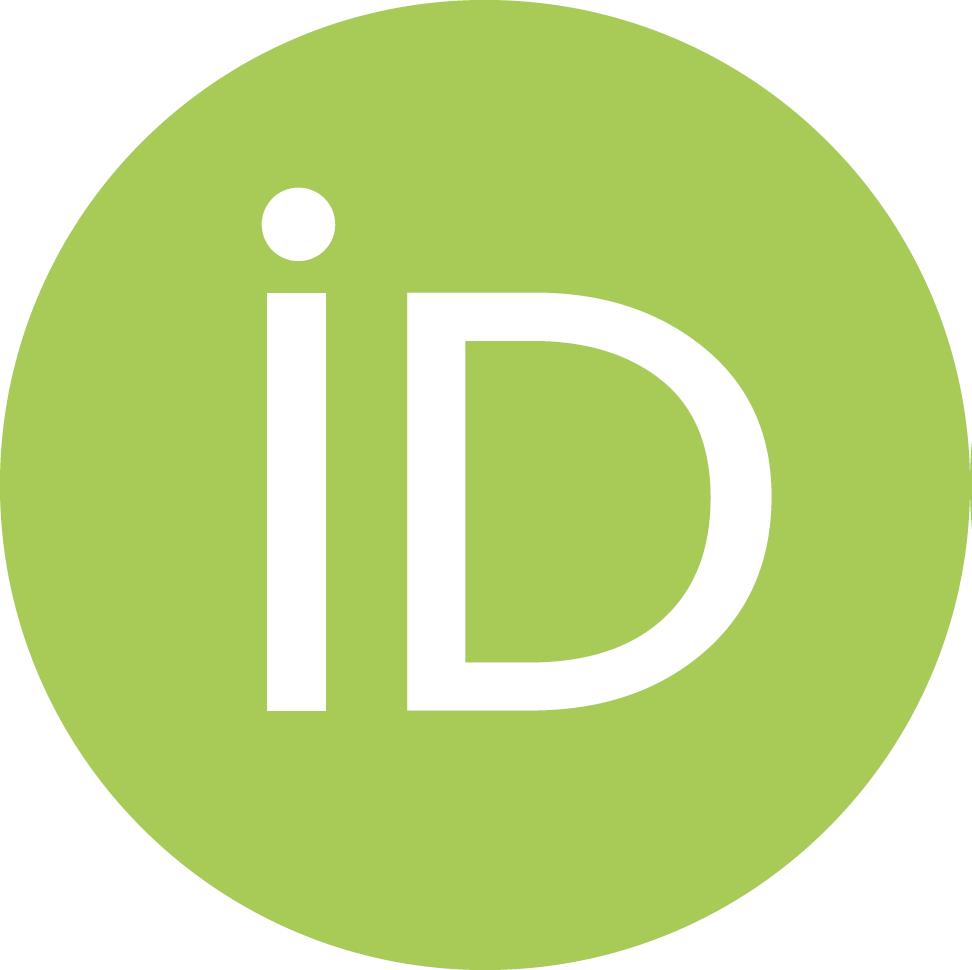}\hspace{1mm}Ruixuan Huang} \\
	National Synchrotron Radiation Laboratory\\
	University of Science and Technology of China\\
	Hefei, Anhui 230026 \\
	\texttt{rxhuang@ustc.edu.cn} \\
	\And
	\href{https://orcid.org/0000-0000-0000-0000}{\includegraphics[scale=0.01]{orcid}\hspace{1mm}Qika Jia} \\
	National Synchrotron Radiation Laboratory\\
	University of Science and Technology of China\\
	Hefei, Anhui 230026 \\
	\texttt{jiaqk@ustc.edu.cn} \\
}

\renewcommand{\shorttitle}{High duty cycle EUV radiation source based on inverse Compton scattering}

\hypersetup{
pdftitle={A template for the arxiv style},
pdfsubject={q-bio.NC, q-bio.QM},
pdfauthor={David S.~Hippocampus, Elias D.~Striatum},
pdfkeywords={First keyword, Second keyword, More},
}

\begin{document}
\maketitle

\begin{abstract}
Inverse Compton scattering (ICS) can obtain quasi-monochromatic and directional EUV radiation via a MeV-scale energy electron beam and a micron-scale wavelength laser beam, which enables a dramatic reduction in dimension and expense of the system, and makes it an attractive technology in research, industry, medicine and homeland security. Here we propose an EUV source based on high repetition ICS system. The scheme exploits the output from the laser-electron interaction between a MW-ps laser at MHz repetition-rate and a high quality electron beam with an energy of a few MeV at MHz repetition-rate.

\end{abstract}


\section{Introduction}

The production of Extreme Ultraviolet radiation (13.5 nm wavelength) is of great significance for advanced lithography technologies. Accelerator-based EUV sources are attracting attention for high power characteristic, such as free electron lasers and synchrotron radiation sources. However, both the size and the cost of such light sources are enormous. One possible way to generate high brightness EUV photons in a more compact and economical device is called inverse Compton scattering (ICS) source~\citep{ICS-Krafft}. ICS can obtain quasi-monochromatic and directional EUV radiation via a MeV-scale energy electron beam and a micron-scale wavelength laser beam, which makes it an attractive technology in research, industry and homeland security.

Recent experiments has demonstrated that ICS source can obtain peak brightness exceeding that of third generation synchrotron light sources (such as~\citep{ICS-Albert}). But the flux yield from a single-shot interaction is limited to 10$^{7}$-10$^{8}$ photons in 1\% bandwidth. In order to meet the practical flux requirements, a high duty cycle ICS system needs to be developed. Specifically, a high repetition-rate low emittance electron beam is required as well as a high repetition-rate high peak power laser beam. Several novel ICS schemes have been recently proposed, including superconducting ERLs for high gamma-ray fluxes, FEL\/ICS hybrid scheme.

Here we propose a high duty cycle EUV radiation scheme based on ICS system, and give a brief view on possible technologies of high repetition rate electron and laser sources.

\section{Inverse Compton scattering}

\subsection{Scattered Photon Energy and Wavelength}

\begin{figure}[!htb]
   \centering
   \includegraphics*[width=.6\columnwidth]{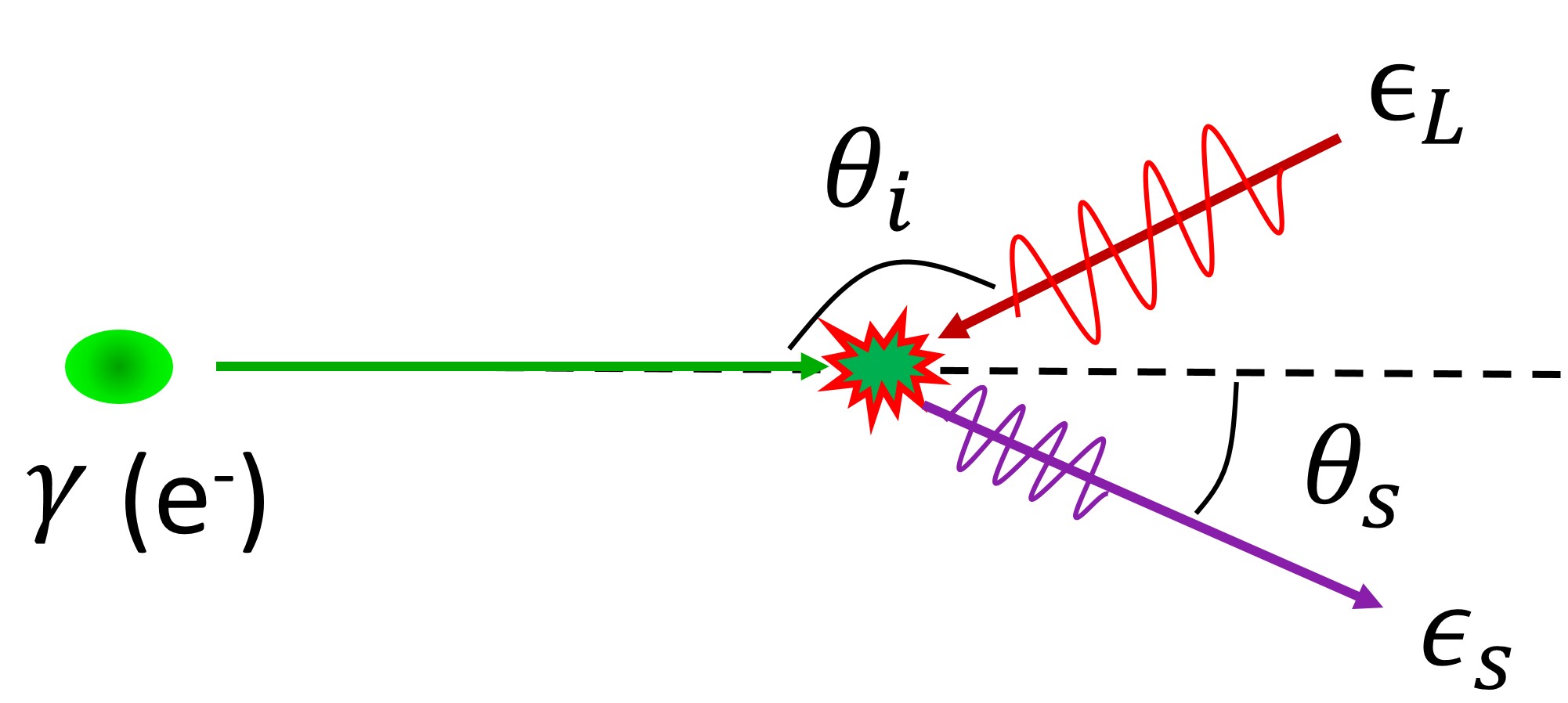}
   \caption{Collision geometry}
   \label{fig:collision}
\end{figure}

ICS is regarded as the elastic collision between electrons and photons. According to energy and momentum conservation, the scattered photon energy $\epsilon_s$ is decided by
\begin{equation}\label{eq:photon_energy}
    \epsilon_s=\frac{(1-\beta\cos\theta_i)\epsilon_L}{1-\beta\cos\theta_s+(1-\cos(\theta_s-\theta_i))\epsilon_L/\gamma mc^2} ,
\end{equation}
where $\epsilon_L$ is the photon energy from incident laser, $\theta_i$ is the interaction angle between the electron and laser beam, $\theta_s$ is the angle between the electron and the scattered photon. $\gamma mc^2$ is the electron energy, and $\beta$ is the relative speed of light, which can be written by the relativistic Lorentz factor $\gamma$ as $\beta=\sqrt{1-1/\gamma^2}$. The collision geometry and angle definitions are shown in Fig.~\ref{fig:collision}.

The scattered photon energy depends on the interaction angle between the laser pulse and the electron pulse, and also the both energies, which gives several options for tuning the wavelength of the light source. If backscattered ($\theta_i=\pi$), the wavelength of the photon is equal to 
\begin{equation}\label{eq:wavelength}
    \lambda_s=\frac{\lambda_L}{4\gamma^2}(1+a^2_L+\gamma^2\theta_s^2) .
\end{equation}

\begin{figure}[!htb]
   \centering
   \includegraphics*[width=.6\columnwidth]{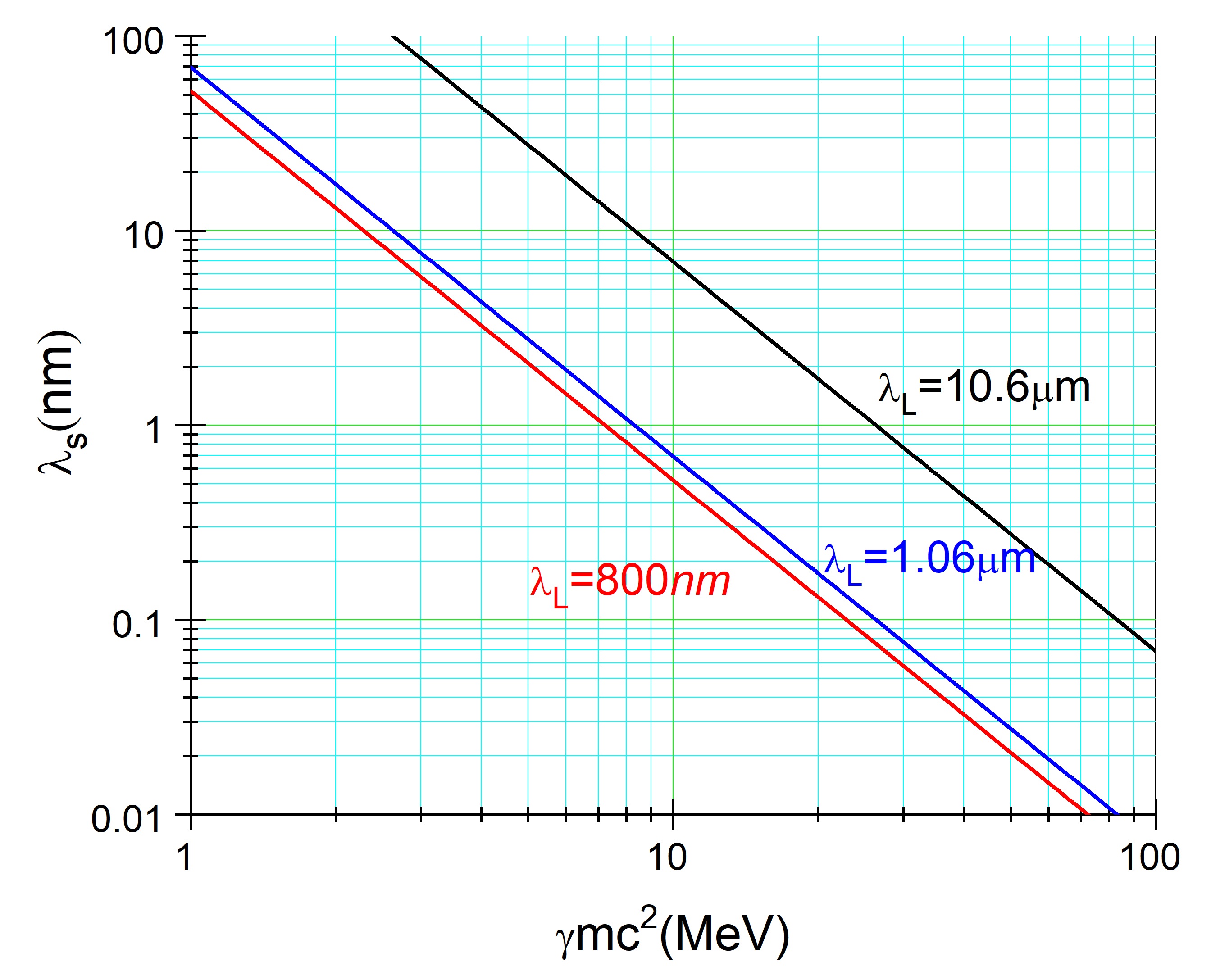}
   \caption{Backscattered photon wavelength as function of electron energy for different laser wavelengths}
   \label{fig:wavelength}
\end{figure}

The incident laser behaves like a virtual undulator with an extremely short period $\lambda_u$ equal to half of its wavelength as $\lambda_u=\lambda_L /2$. The equivalent undulator parameter is $a_L^2=4\gamma\epsilon_L/mc^2$. A relatively short wavelength of laser, normally in the range of 0.3--10.6 $\mu$m, allows relatively easy access to the EUV and X-ray region. Figure~\ref
{fig:wavelength} presents the relationship between the photon wavelength and the electron energy for different incident laser wavelengths~\citep{Jia}. To produce EUV photons of 13.2 nm wavelength, an IR laser of 800 nm wavelength with an energy beam of 2 MeV energy, or an IR laser of 10.6 $\mu$m wavelength with an energy beam of 7 MeV energy is required.

It is notable that the scattered photon energy is maximal in the electron beam direction, and the half angle of the emission cone is in the order of $1/\gamma$. Thus a higher beam energy can expect a narrower emission cone therefore a higher photon brilliance.

\subsection{Radiation power}

The number of the emitted photons per second is determined by
\begin{equation}\label{eq:number}
    N_s=\frac{N_e N_L f_s\sigma_C}{\pi (\sigma_e^2+\sigma_L^2)} ,
\end{equation}
where $N_e$ and $N_L$ are the numbers of the electron and incident laser photon in one pulse, respectively. The $\sigma_C$ represents the cross section of the Compton scattering, $f_s$ is the collision frequency which is decided together by the repetition rates of the electron and the laser beam. $\sigma_e$ and $\sigma_L$ are the rms spot sizes of the electron and the laser beams, respectively. If the polarization of the scattered photons is not concerned, the cross section is given by
\begin{equation}\label{eq:cross_section}
    \sigma_C=(8\pi/3)r_0^2 \approx 6.65 \times 10^{-29} \mathrm{m^2} ,
\end{equation}
where $r_0$ is the classical electron radius. Since the interaction has a significantly low cross section, actions should be taken to increase the total photon flux. Effective ways include increasing the electron bunch charge and the laser pulse energy, focusing the beam sizes at the collision point, and operating the system at a higher duty cycle.

For a laser beam with an energy of $E_L$, the photon number in a laser beam is given by $N_L=E_L\lambda_L/(hc)$, where $\lambda_L$ is the wavelength and $h$ is Plank constant. The average radiation power $P_\mathrm{avg}=N_s(hc/\lambda_s)$ has the engineering formula of
\begin{equation}\label{eq:average_power}
    P_{avg} (\mathrm{mW}) \approx 0.1323 \frac{\lambda_L(\mu\mathrm{m})}{\lambda_s (\mathrm{nm})} \cdot \frac{ Q_e (\mathrm{pC}) E_L (\mathrm{J}) f_s (\mathrm{kHz}) } { \sigma_e^2 (\mu\mathrm{m} ) + \sigma_L^2 (\mu\mathrm{m}) },
\end{equation}
where $Q_e=N_e e$ is the electron bunch charge. The duration of the emitted photon $\tau_s$ is the pulse stretching due to the photon and electron speed mismatch over the interaction time as $\tau_s \approx \tau_e+\frac{\tau_L}{4\gamma^2}$, where $\tau_e$ and $\tau_L$ are durations of the electron bunch and laser pulse, respectively. Practically it has $\tau \ll 4\gamma^2 \tau_e$, we can consider $\tau_s \approx \tau_e$. Then, the radiation peak power is determined by the number of the photons and the pulse duration, which has the engineering formula of
\begin{equation}\label{eq:peak_power}
    P_{peak} (\mathrm{kW}) \approx 132.3 \frac{\lambda_L(\mu\mathrm{m})}{\lambda_s (\mathrm{nm})} \cdot \frac{ Q_e (\mathrm{pC}) E_L (\mathrm{J})  /\tau_e (\mathrm{ps}) } { \sigma_e^2 (\mu\mathrm{m} ) + \sigma_L^2 (\mu\mathrm{m}) }.
\end{equation}
For example, at $E_L$=1~J, $\lambda_L$=10.5$\mu$m, $Q_e$=300~pC, $\sigma_e = \sigma_L$ = 50 $\mu$m, $\tau_e$ =2 ps, the EUV light with an energy of 92 eV would have 4.2$\times 10^8$ photons per bunch, and the desirable peak power could reach 3 kW.

The quality of the generated EUV beam depends heavily on the characteristics of the electron and laser beams involved in the interaction. The relative bandwidth of emitted radiation $\sigma_{E_s}/{E_s}$ is jointly decided by the aperture energy spread $\sigma_\gamma / E_\gamma$, laser energy spread $\sigma_L / E_L$, and the beam emittance $\sigma_\varepsilon/E_\varepsilon$. In the Thomson regime where electron beam energy is relatively low, the beam emittance dominates.

\section{High repetition rate electron guns}

High duty cycle ICS sources desire high repetition rate electron beams with characteristics very similar to those of electron sources required in X-ray FELs. It is natural to use a photocathode-based injector due to its advantage of low emittance. Delivering the required beam current requires high quantum efficiency (QE) cathodes such as semiconductors. The state-of-art S-band photocathode gun typically operates at or below 100 Hz repetition rate because of the heat load on the cavity walls. Many groups over the world are working on electron sources at MHz/GHz repetition rates. Some technologies already provide sources with quality matching the high duty cycle EUV ICS sources. The list includes DC guns, SRF guns and VHF guns.

\subsection{DC Guns}

High voltage DC schemes allow for arbitrary high repetition rates, which have demonstrated the high average current and low transverse emittance with photoemission cathodes. A typical performance of a DC gun is to produce a several tens of pC bunch charge with several $\mu$m  emittance at hundreds of MHz or GHz repetition rate. Since the DC guns are approaching the limit of the cathode gradients and beam energies, it is difficult to achieve a significant improvement on beam quality.

The best performing DC photoemission guns are the FEL gun and polarized electron sources at JLab, and the ERL gun at Cornell. The Cornell DC gun operated at 350 kV can deliver an average current up to 100 mA by $\sim$77 pC bunches at 1.3 GHz repetition rate~\citep{Cornell-Gun}. Recently low emittance measurements performed with a NaKSb cathode illuminated by a 50 MHz laser have demonstrated capacity of high quality beam with 0.78 $\mu$m normalized emittance at 300 pC charge per bunch. 

\subsection{SRF Guns}

Electron guns based on SRF schemes allow for very high repetition rates (GHz-class) and have the prospect to achieve energy goals. Thanks to the effective cryo-pumping from the superconducting walls, they show an excellent vacuum performance. The main challenges are to find a suitable semiconductor cathode under the SRF environment, to overcome the degradation on the peak field and QE lifetime of cathode and to solve the compatibility issues of beam focusing due to field exclusion.

The 3.5-cell 1.3 GHz Rossendorf SRF gun has been operating for over 10 years as an injector of the ELBE infrared FEL. The Cs$_2$Te  photocathode with a QE of 1\% has been proved the good compatibility with the SRF environment. The design parameters of the Rossendorf gun are the electron energy of 9.5 MeV, bunch charge of 77 pC at the repetition rate of 13 MHz with the transverse emittance of 1.0 $\mu$m. Several upgrades are underway including the cavity design improvement and the testing of transverse focusing by a superconducting solenoid or the excited high order modes in the gun cavity. BNL is developing several SRF guns. The 700 MHz half-cell SRF gun is designed for ERL applications to accelerate the beam up to $\sim$2 MeV~\citep{SRF-Gun}. The high current mode expects to provide 500 pC bunches in CW.

\subsection{VHF Guns}

Normal conducting RF gun, such as S-band or L-band gun, can fulfill the beam requirements by the single-pulse ICS source, while has a fairly low duty cycle. By decreasing the operating frequency, the cavity size increases with a significant reduce of the power density on the structure wall, which contributes to a higher duty cycle. The Boeing gun has achieved 25\% duty cycle operating at 433 MHz. The LANL 700 MHz RF gun has realized a CW operation mode with a complicated cooling system. The LBNL developed a VHF gun operating at CW 186 MHz~\citep{VHF-Gun}. The nominal beam energy at the gun exit is 750 keV with an acceptable decrease of the peak field (19.5 MV/m) at the cathode. The VHF gun at APEX injector facility has demonstrated all the beam quality required by high repetition rate X-ray FELs. 

Recently, a two-cell version of VHF gun called APEX-2 was proposed with the fields and beam energy comparable to those targeted by SRF guns but with a lower cost and complexity. The gun can be CW operated at 162.5 MHz (the 1/8 sub-harmonics of 1.3 GHz) with an enhanced launching field (34 MV/m) at the cathode~\citep{APEX2}. Initial simulation indicates that electron beam with 0.13 $\mu$m emittance at 100 pC bunch charge and 2 MeV energy could be pursued. Such a VHF gun can not only represent a low risk and cost-effective way of achieving the energy and emittance required by high duty cycle ICS source, but also provide a natural upgrade to CW FEL system with is a very promising candidate for a high throughput EUV lithography.

\section{High average power laser system}

The laser system is a key portion of the ICS source, it should guide both the cathode drive laser and the Compton scattering. To produce a high flux EUV source, a high power and tightly focused laser pulse with a repetition rate that matches the electron beam is required. However, an IR laser with several Joule pulse energy and ps-MHz pulse structure laser is not commercially available. Instead, IR laser with advanced amplification technology~\citep{EnhanceCavity} is required to enhance the laser power at collision. It is estimated that a MW-ps laser at MHz repetition-rate  is demanded for a mW average power EUV ICS source.

\section{Conclusion}
We have discussed the EUV source based on high repetition rate ICS system, and given a brief view on possible technologies of high repetition rate electron and laser sources. The upgraded VHF-gun technology and the IR laser with an advanced amplification technology are suggested. Theproposed compact ICS scheme is efficient and may be used for EUV lithography and other potential applications.

\bibliographystyle{unsrtnat}
\bibliography{references}  

\end{document}